# *Logical inconsistency in quantum mechanics*

B. E. Y. Svensson

*Theoretical High Energy Physics,*
*Department of Astronomy and Theoretical Physics,*
*Lund University, Sölvegatan 14A, SE-22362 Lund, Sweden*

Abstract

I show that the application of the quantum-mechanical (QM) "which-way" weak measurement scheme of Vaidman may lead to logical inconsistencies. To this end, I study weak values of projection operators. Weak values are (normalized) amplitudes, operationally defined by a weak measurement followed by postselection. Projector weak values have a direct physical significance. This allows formulating an inconsistency in classical terms, *viz.*, the contradiction in the readings of several measuring devices. To arrive at the contradiction, I also invoke the property of weak measurements not to "collapse the wave function" but to leave the state of the system unchanged (to lowest order in the weak measurement strength). My arguments rely entirely on basic QM rules plus commonly accepted weak QM measurement approximations. Therefore, the inconsistency challenges QM as such.

1. **Introduction**

Famously, in his lectures [1] Feynman pointed out how the basic rules of quantum mechanics (QM) may lead to "quantum weirdness". By this is meant that QM sometimes leads to results that, based on classical preconceptions, may seem perplexing. These "weird" results, to the extent that they have been tested experimentally, have up until now been brilliantly confirmed.

The review article [2] digs deeper into the questions how well quantum phenomena can be understood in classical terms. Among other things, this article exhibits situations in which only a genuine QM description applies, *i.e.*, where no classical description whatsoever is able to reproduce the QM findings.

It should be underlined that "weirdness" does not mean logical inconsistency. To arrive at a logical inconsistency one must exhibit a logical contradiction, *i.e.*, one must show that a statement *p* together with its negation ¬ *p* both hold true. Whether this can be exhibited for QM partly depends on exactly what QM postulates one invokes. For example, the authors of [3] have arrived at a no-go theorem for QM, assuming a unitary evolution for *all* systems, even macroscopic ones, *i.e.*, not assuming any von Neumann-like [4] "collapse of the wave function".

In this article, I investigate QM from a different point of view than in [3]. I do assume standard, orthodox QM, not excluding the von Neumann collapse postulate. What I do use is the concept of weak measurement, combined with the notion of postselection leading to the concept of a weak value [5] (see [6 - 7] for some reviews). More specifically, I use the fact that a weak value is a (normalized) transition *amplitude*, implying that a weak measurements allow one to study the amplitudes themselves. In a previous paper [8], I investigated whether the use of weak values of (combinations of) projection operators was free of inconsistencies and found examples where this was not the case. In the present article, I extend this analysis and argue that the inconsistencies reach the core of QM as such.

My line of attack is strongly inspired by Vaidman's approach [9 - 10] to "the past of a quantum particle". Indeed, the argument I put forward is based on nothing but a slight reformulation of his approach. However, my presentation is arguably somewhat more basic than what, to my knowledge, up untill now exists in the literature.

I stress from the outset that my treatment is based on nothing but ordinary QM and its standard interpretation, including the weak limit approximations commonly accepted in deductions within the QM weak measurement approach.

In the next section, I present the basis of my approach. In the following sections I point out how a projector weak value may be used to establish the presence or not of an intermediate state in the QM description of a system, and apply it to a pair of projectors to arrive at a logical contradiction. A final section gives my conclusions. In the appendix, I illustrate my approach and conclusions by the so called three-box arrangement [11-12].

2. **Basis of the "which-way" approach**.

Vaidman [9 - 10] was interested in studying the evolution of a quantum system from its preparation in an initial, "preselected" state $|in>$ at a time $t_i$ until its detection at a later time $t_f$ in a final, "postselected" state $|f>$. The relevant entity for investigation of such evolution is the transition amplitude $<f|U(t_f, t_i)|in>$, where $U(t_f, t_i)$ is the unitary operator for the evolution from $t_i$ to $t_f$ of the system under study. In particular, Vaidman proposed a procedure for answering the question: could the system, at an intermediate time $t_m$, $t_i < t_m < t_i$, have occupied a specific eigenstate $|a_o>$ of a given system operator $A$?

The first step of that procedure is to insert a complete set of (orthogonal) eigenstates $\{|a_k>\}_{k=0}^{k=N}$ of the operator $A$ – here assumed to have a discrete, finite and non-degenerate spectrum – into the transition amplitude at time $t_m$:

$$<f | U(t_f, t_i) | in> = \Sigma_k <f | U(t_f, t_m) | a_k><a_k | U(t_m, t_i) | in>. \quad (1)$$

Concrete examples of such intermediate states –which I also call "channels" – are the different paths that a photon may take through a set of interlocked Mach-Zehnder interferometers (MZIs) [10], the boxes in the three-box arrangement [11-12] or the different "pigeon-holes" that the system may occupy in the quantum pigeon-hole setup [13].

Note also that formula (1) expresses the Feynman dictum [1] of arriving at the transition probability $prob(in \rightarrow f) = |<f | U(t_f, t_i) | in>|^2$ by summing over all (indistinguishable) intermediate states before taking the square modulus. Indeed, the fact that QM probabilities are obtained from *square moduli* of transition amplitudes is vital for QM consistency, *i.e.*, for "weirdness" not to imply logical contradictions.

The second step in the procedure is to attach a von Neumann measuring device to the system at the "channel" $| a_o >$, weakly measuring the projection operator $\Pi_o = | a_o ><a_o |$ at the intermediate time $t_m$. In the usual treatment, this involves taking the limit $g \rightarrow 0$ of a transition operator $exp [-i g \Pi_o \otimes P_M]$ describing an (impulsive) interaction between the system and a "von Neumann meter" with "pointer variable" canonical coordinates $Q_M$ and $P_M$.

Two characteristics of this weak measurement procedure are of particular importance for my further arguments.

The first is that, to zeroth order in the strength $g$ of the weak measurement, the state $U(t_m, t_i) | in >$ of the system just before the weak measurement remains the same immediately after that measurement: in the weak measurement limit, there is no "collapse of the wave function" [6]. I shall, moreover, specialize to the case of the system undergoing no non-trivial internal time-evolution during some finite time interval $\Delta t$ after $t_m$; this assumption amounts to assuming that the system's state remains unchanged (to lowest order in $g$) during this time interval. During $\Delta t$ one may thus perform weak measurements of several different system projectors with the system represented (to lowest order in $g$) by the same state $U(t_m, t_i) | in >$ as at time $t_m$.

The second characteristic of the weak measurement procedure is the role of the meter [6]. Namely, the mean value $<Q_M>_f$ of the pointer variable $Q_M$, after postselecting the state $|f>$, supplies the so called weak value $(\Pi_o)_w$ of $\Pi_o$:

$$lim_{g \rightarrow 0}(<Q_M>_f / g) = <f | U(t_f, t_m) \Pi_o U(t_m, t_i) | in > / <f | U(t_f, t_i) | in > \equiv$$

$$\equiv <f | U(t_f, t_m) | a_o><a_o | U(t_m, t_i) | in > / <f | U(t_f, t_i) | in > \stackrel{\text{def}}{=} (\Pi_o)_w . \quad (2)$$

(For simplicity, I here assume the pointer variable mean value $<Q_M>_f$ to be zero for the idle meter. I also assume the value $(\Pi_o)_w$ to be real; to get at an imaginary part of $(\Pi_o)_w$ needs measuring $<P_M>_f$ in the same limit. Furthermore, I assume that the denominator $<f | U(t_f, t_i) | in >$ of $(\Pi_o)_w$ does not vanish.)

The weak value $(\Pi_o)_w$ is thus nothing but – in the numerator –the transition amplitude $<f | U(t_f, t_m) \Pi_o U(t_m, t_i) | in>$ through the state $| a_o >$ of Feynman's formula (1), normalized – in the denominator – by the total transition amplitude $< f | U(t_f, t_i) | in >$.

The weak value of a projector is thus a measurable quantity, albeit requiring repeated identical experiments to get at the statistically defined quantity $< Q_M >_f$. Moreover, not only is it measurable, it is also amenable to a physical interpretation., as I now show.

### 3. The physical interpretation of a projector weak value

The entity $(\Pi_o)_w$ differs from zero if and only if the system on its evolution to the final state $|f>$ could occupy the state $| a_o >$ at time $t_m$, in the sense that there is a non-zero contribution to the sum in eq.(1) from the term involving the state $| a_o >$. In other words, the non-vanishing or not of $(\Pi_o)_w$ – i.e. whether the von Neumann meter gives a signal or remains idle – supplies a YES/NO answer to the question: is the state $| a_o >$ a possible intermediate state of the system at time $t_m$ ? (Some intricacies are involved here, though, related to what in [14 - 16] is called "non-representative weak values" of projectors. I assume throughout the present article that all weak values are "representative" in the sense mentioned there.) And this answer is provided in terms of *classical* pieces of data, *viz*., the triggering or not of the von Neumann meter.

I stress that it is the non-vanishing or not of the weak value that is of importance for the triggering of the meter; whether the weak value is positive or negative is of no importance (nor is the possibility of it taking a complex value, provided one were to allow for a meter also measuring $< P_M >_f$). Neither do any considerations of "contextuality" (see, *e.g.*, [17-18] and references therein) or the like enter.

### 4. Comparison between weak and strong measurements.

Note the similarities but also the differences between weak measurements and ordinary, strong or projective measurements.

A strong measurement of a projector provides a (probabilistic) answer to whether the state $| a_o >$ is a possible intermediate state of the system. Moreover, QM furnishes the probability $prob(in \to a_o) = | < a_o| U(t_m, t_i) | in >|^2$ for the outcome of such measurement (or, if a postselection applies, the probability $prob (f / a_o, in) \equiv prob ( in \to f\ via\ a_o ) = |<f | U(t_f, t_m) | a_o > < a_o| U(t_m, t_i) | in >|^2$ ). A strong measurement thus furnishes a classical piece of data in form of this probability.

Furthermore, the state of the system after a strong measurement with outcome $| a_o >$ "collapses" to that state, implying that any immediately following measurement on the system will be on this collapsed state, not on the unperturbed system state.

On the other hand, within standard QM, weak projector measurements can directly offer only a YES/NO answer to whether the state $| a_o >$ is a possible intermediate state of the system. (Since the absolute square of the numerator of the weak value $(\Pi_o)_w$ equals $prob(in \rightarrow f \ via \ a_o)$, weak measurements of both real and imaginary parts of $\Pi_o$, plus a measurement of $prob \ (in \rightarrow f) \equiv prob \ (f \ / \ in) = |<f| \ U(t_f, t_i)| \ in>|^2$, could indeed supply the probability $prob \ (f \ / \ a_o, \ in)$. It would, however, be in a more indirect way, requiring several different kinds of measurement.) This answer is provided by the triggering or not of the von Neumann meter, *i.e.*, in terms of *classical* data. In other words, a weak measurement can directly tell whether, at the intermediate time $t_m$, the total state of the system on its evolution to the postselected state, could proceed or not through the state $| a_o >$.

What is even more important, a weak measurement, contrary to a strong one, does not "collapse" the state but leaves it unperturbed, to leading order in the weak measurement strength $g$ [6].

In short, a weak projector measurement directly tests for the presence of an *amplitude* without (to lowest order) disturbing the system. A strong measurement, however, measures a corresponding *amplitude squared*, at the same time in general heavily disturbing the state of the system.

Note in particular that the question regarding the presence of a particular state in the expansion (1) of the system's total transition amplitude – a statement *within* the *QM* formalism – has, through the weak measurement followed by postselection, been reduced to a registration (albeit a statistical one) in a measuring apparatus, *i.e.*, a *classical* datum.

## 5. Application to a pair of projectors.

Let me now apply this general framework to two different projectors for the system, say $\Pi_o = | a_o > < a_o |$ and $\Pi_1 = | a_1 > < a_1 |$, both being projectors onto eigenstates of the given operator $A$. Since I assume $| a_o >$ and $| a_1 >$ to be orthogonal, the combination $\Pi^+ = \Pi_o + \Pi_1$ is also a projector. The projector $\Pi^+$ could be a non-local operator, for example involving intermediate states representing different spatial locations in a nested MZI. The measurement of such non-local projectors should, however, cause no problems of principles regarding simultaneous measurements within the framework of non-relativistic QM; such measurements are (at least implicitly) assumed, *e.g.*, in [11].

Next, think of weakly measuring $\Pi^+$, $\Pi_o$ and $\Pi_1$ separately by separate von Neumann measuring devises, each with their own measurement strength. These measurements are thought of as taking place during the time interval $\Delta t$ defined above, *i.e.*, during which the state of the system remains unchanged (to lowest order in the weak measurement strengths). Then $(\Pi_o)_w$ tests for the presence of $| a_o >$ as an intermediate state, and $(\Pi_1)_w$ independently tests for the presence of $| a_1 >$. The weak value $(\Pi^+)_w$ tests for the presence of either $| a_o >$ or $| a_1 >$ (or both): if the $\Pi^+$-meter triggers, one may conclude that either $| a_o >$ or $| a_1 >$ (or both) are possible intermediate states, if it does not trigger one concludes that neither is a possible intermediate state.

### 6. The contradiction.

Now comes the punch line.

It is easy to find even simple systems (see the appendix) for which $(\Pi^+)_w$ vanishes, while both $(\Pi_o)_w$ and $(\Pi_1)_w$ individually are different from zero. From what has been said, this is interpreted to mean that both $|a_o>$ and $|a_1>$ are present as intermediate states when tested individually, while none is present when tested as a pair. Since the vanishing or not of the weak values involved are *classical* pieces of data – registrations in the respective von Neumann meters – this is a logical contradiction.

In trying to understand the basic cause for this contradiction, one must keep in mind that a weak value is a (normalized) *amplitude* while at the same time providing a *classical* piece of data through the weak von Neumann protocol. And while *interference between amplitudes* is legion, it is crucial in the usual arguments for the logical consistency of QM that its basic rules furnish *classical data* in terms of *amplitudes squared*. Taking the square modulus – and observing that the "collapse of the wave function" has decisive influence on sequential measurements – so to speak "hides" the consequences of such interferences. As this usual argument goes, there are no logical inconsistencies, at most "merely" the well-known QM "weirdnesses" in terms of clashes between QM predictions and preconceived, classically based expectations [1-2].

On the other hand, invoking weak measurements and postselection allows one, metaphorically speaking, to "sneak inside the square modulus" to study intermediate transition *amplitudes* directly, at the same time being sure that the measurements are made on the same state of the system. Moreover, one may establish the vanishing or not of these amplitudes in terms of *classical* data– the triggering or not of a meter. This enables the direct study of the consequences of interference between amplitudes in terms of classical concepts, leading to the type of logical contradictions pointed out here.

### 7. Conclusion

Using standard quantum mechanics applied to weak measurements of projectors, I have derived a logical contradiction expressed in terms of three classically phrased statements. Since my arguments are built on nothing but basic quantum mechanics and commonly accepted weak measurement approximations within that same framework, I see no escape from the conclusion that these inconsistencies have their roots in the fundaments of quantum mechanics.

**Acknowledgement**

I thank J. Bijnens, R.B. Griffiths, R. Renner, M. Sjödahl, B. Söderberg and, particularly, E. Cohen for valuable comments to preliminary versions of my article.

### Appendix

The so-called three-box arrangement [11-12], furnishes a simple example of the situation treated in the main text.

Consider a one-particle system, with no non-trivial time evolution, in which the particle could be in any of three "boxes" $A$, $B$ or $C$, represented by states $|A>$, $|B>$ and $|C>$ in a three-dimensional Hilbert space. Let the preselected state be

$$|in> = (|A> + |B> + |C>)/\sqrt{3} \;, \qquad (A1)$$

and the postselected state be

$$|f> = (|A> + |B> - |C>)/\sqrt{3} \;. \qquad (A2)$$

One is interested in finding which intermediate states the particle may have occupied, *i.e.* properties of the projectors $\Pi_A = |A><A|$, $\Pi_B = |B><B|$ and $\Pi_C = |C><C|$ at an intermediate time.

One easily finds both weak values $(\Pi_A)_w$ and $(\Pi_C)_w$ to be non-zero, a fact that is interpreted as the presence of $|A>$ and $|C>$ as possible intermediate state. But testing this on $\Pi_A + \Pi_C$ gives a vanishing value for $(\Pi_A + \Pi_C)_w$, so no particle intermediately neither in box $A$ nor in box $C$.